\documentclass[%
 reprint,
 amsmath,amssymb,
 aps,
]{revtex4-1}

\usepackage{graphicx}
\usepackage{dcolumn}
\usepackage{bm}

\begin{document}

\title{
\hspace*{13cm}{INR-TH-2014-002}  \\
\begin{center}
Multi-particle Processes and Tamed Ultraviolet Divergences
\end{center}}

\author{Victor~T.~Kim}
\email{kim@pnpi.spb.ru}
\affiliation{St. Petersburg Nuclear Physics Institute NRC KI, Gatchina 188300, Russia \\
St. Petersburg State Polytechnical University, St. Petersburg 195251, Russia}
\author{Grigorii~B.~Pivovarov}
 \email{gbpivo@ms2.inr.ac.ru}
 \affiliation{Institute for Nuclear Research of RAS, Moscow 117312, Russia}




\date{\today}

\begin{abstract}
New approach to computing the amplitudes of multi-particle processes in renormalizable quantum field theories is presented. Its major feature is a separation of the renormalization from the computation.
Within the suggested approach new computational rules are formulated. According to the new rules, the amplitudes under computation are expressed as a sum of effective Feynman amplitudes whose vertexes are the complete amplitudes of the processes involving not more than four particles, and the lines are the complete two-point functions. The new rules include prescriptions for computing the combinatorial factors by each amplitude. It is demonstrated that due to these prescriptions the combinatorial factors by the amplitudes that are divergent in the ultraviolet in four space-time dimensions vanish. Because of this, the computations within the new approach do not involve the ultraviolet renormalization. 
\end{abstract}

\maketitle


\section{Introduction}

Physics at the LHC is predominantly the physics of multi-particle processes. Substantial efforts have been invested in developing the techniques for computing the amplitudes of such processes (see, e.g., \cite{Kilian:2007gr} and \cite{Binoth:2010ra}). One of the difficulties in organizing such computations is in keeping track of all the necessary renormalizations.

In this paper we present a new approach for computing the multi-particle amplitudes. Its major feature is a separation of the renormalization from the computation. The idea is that the renormalization is needed only for computing the basic processes involving not more than four particles. After computing the basic amplitudes, it should be possible to compute the multi-particle amplitudes combining the basic ones, and this combination should not involve any renormalization.

Let us compare the new approach to the well-known approach of effective action \cite{Vasiliev:1998cq}. Within the effective action approach, the connected amplitudes are constructed not from the bare action of the model, but from the effective action, which is the generating functional of the one-particle irreducible amplitudes. 

The effective action involves nonlocal effective vertexes, and the connected amplitudes are composed from the effective vertexes with tree graphs. These tree graphs are generated with the inversion of the Legendre transform. The edges of the graphs correspond to exact propagators. There are infinite sequence of the effective vertexes ordered by the number of external legs of the corresponding one-particle irreducible amplitudes. The UV renormalization is needed only for computing the effective vertexes and the exact propagators.

In contrast, within the new approach, the connected multi-particle amplitudes are composed not from the bare action, or from the effective action, but from a nonlocal functional of fields which is the generating functional of connected amplitudes with the number of external legs bounded from above. The exact form of this bound is related to properties of the bare action of the model. In the case of the renormalizable field theories in four space-time dimensions, the connected amplitudes used as effective vertexes involve not more than four external legs.

To stress the difference between the new approach and the approaches employing various actions, we call it the inaction approach, and the generating functional of the effective nonlocal vertexes of the new approach, the inaction functional. The inaction functional is a nonlocal polynomial of the fields.

Like in the effective action approach, the UV renormalization is needed within the inaction approach only for computing the inaction functional and the exact propagators. Also, like in the effective action approach, there is a procedure for composing the multi-particle connected amplitudes from the vertexes of the inaction functional and the exact propagator. This procedure is parallel to the inverse Legendre transform of the effective action approach. We call it the quantum transform. A crucial difference between the  effective action and the inaction approaches is that the quantum transform generates not only the tree graphs, but also graphs with loops. Therefore non-trivial momentum integrations appear as an outcome of the quantum transform.

The above comparison can be summarized as follows: the inaction functional is simpler than the effective action, but the quantum transform is more involved than the inverse Legendre transform. In comparison with the bona fide perturbation theory based on the bare action, the nonlocal inaction functional is more involved than the bare action, but the quantum transform is in some sense (see below) simpler than the conventional Feynman rules. So, the inaction approach lies somewhere in between the bona fide perturbation theory and the effective action perturbation theory.  

Now we describe the general idea behind the quantum transform. The starting point in deriving the quantum transform is the Dominicis-Englert duality between the bare action and the generating functional of connected Green's functions \cite{deDominicis:1967px}. In particular, this duality implies that if the generating functional $W(J)$ of connected Green's functions is given, the bare action functional $S(\phi)$ correponding to this $W(J)$ can be expressed in terms of $W(J)$ (here $\phi$ denotes the multicomponent field of the model, and $J$, its source).

Next, one can express restrictions on the models one considers in the form of the equation $PS(\phi)=0$, where $P$ is a projector acting in the linear space of field functionals. Substituting in this equation the expression of $S(\phi)$ in terms of $W(J)$ one obtains what we call \emph{an inaction equation} for connected amplitudes.  

The inaction equation can be solved for the part of the generating functional of connected amplitudes in the range of the projector $P$. The part of this generating functional in the kernel of $P$ is used to parameterise the theory. We called it above the inaction functional. Within the inaction approach, the inaction functional can vary without restrictions inside the kernel of $P$. The solution yields the part of the generating functional of connected amplitudes in the range of $P$ as the above quantum transform of the inaction functional.

The quantum transform depends on the projector $P$. Requiring the physical results to be independent of the choice of $P$ leads to a particular formulation of the renormalization group equations.

The inaction approach was suggested in \cite{Pivovarov:2009wa}.
In that paper, the approach was considered for $\phi^4$. In particular, it was demonstrated that the evolution  of the scalar mass described with the inaction renormalization group equation develops a Landau pole in the running scalar mass.

In \cite{Pivovarov:2012wz} the inaction formulation was extended to gauge theories. In unpublished notes \cite{Pivovarov}, a possibility to use the inaction approach beyond perturbation theory is discussed.

In this paper we consider a particular version of the inaction approach geared to description of multi-particle processes. It results from a particular choice of the projector annihilating the bare action of the theory. Namely, in this paper, the projector nullifies all powers of the fields up to the fourth power inclusively. So, the inaction equation $PS(\phi)=0$ is satisfied for any model which bare action is a polynomial of the fields of the fourth power.

Accordingly, the inaction functional is a nonlocal polynomial of the fields of the fourth power, and the quantum transform expresses the connected amplitudes involving more than four particles in terms of the inaction functional. 

The quantum transform in this case can be represented as a sum of Feynman amplitudes whose propagators are the exact propagators of the model, and the
vertexes are the connected amplitudes with not more than four external legs. It implies certain rules for computing the combinatorial weight with which an amplitude contributes to the transform.

As mentioned above, loop amplitudes appear in the quantum transform. By construction, the quantum transform is UV finite for the exact propagator and inaction functional corresponding to a renormalizble theory. Practically, it is of importance to know the mechanism by which this UV finiteness is maintained.

Surprisingly, the simplest possible mechanism of this finiteness seems to be in action here: The combinatorial weights for UV dangerous amplitudes which we were able to check are simply zero.

We stress that this fact is not proven as of this moment. Below we detail the rules for computing the combinatorial weights appearing in the quantum transform, and give examples of disappearance of some UV dangerous amplitudes. 

Assuming that this convenient feature holds generally, we have the following rules for computing the multi-particle amplitudes: First, the propagators and the amplitudes involving not more than four particles are computed with the standard methods up to desired accuracy. Next, the later are taken to be the vertexes, and the former, the lines of the effective Feynman rules for computing the multi-particle amplitudes. Last, all the UV divergent amplitudes are thrown away. 

We conclude that the quantum transform is simpler than the Feynman rules, because it generates less amplitudes (the UV divergent amplitudes do not appear).

The paper is organized as follows. In the next section, we describe the quantum transform. In the third section, we describe the combinatorial rules for computing the weights of individual Feynman amplitudes in the quantum transform. In the fourth section, we give an example of disappearance of the UV dangerous amplitudes in the quantum transform. In the last section, we discuss and summarize the results. 

\section{The Quantum Transform}
\label{quantum}

The exponential of the generating functional of connected Green's functions $W(J)$ is expressed in terms of the bare action $S(\phi)$ by the functional Fourier transform:
\begin{equation}
\label{FT}
e^{W(J)}=N\int\mathcal{D}\phi\,e^{-S(\phi)+iJ\phi}.
\end{equation}
Here $N$ is the normalization defined by the condition $W(0)=0$.
$S(\phi)$ is UV divergent, while $W(J)$ is UV finite.

Then the exponential of the bare action is expressed in terms of $W(J)$ with the inverse Fourier transform:
\begin{equation}
\label{IFT}
e^{-S(\phi)}=N'\int\mathcal{D}J\,e^{W(J)-iJ\phi},
\end{equation}
where $N'$ is another normalization factor which we do not need to specify.

We assume that $J$ is the source for the deviation of the field from the vacuum value, and, consequently, there is no term linear in $J$
in the expansion of $W(J)$ in powers of the source:
\begin{equation}
\label{bar}
W(J)=-\frac{1}{2}JDJ + \bar{W}(J).
\end{equation}
Here $D$ is the matrix of exact propagators, and $\bar{W}$ is the generating functional of connected Green's functions with more than two external legs.

With the standard functional tricks (see \cite{Pivovarov:2009wa}), we rewrite the right hand side of Eq. (\ref{IFT}) as follows:
\begin{equation}
\label{DE}
e^{-S(\phi)}=\bar{N}e^{-\frac{1}{2}\phi D^{-1}\phi}T e^{A(\phi)}.
\end{equation}
Here $\bar{N}$ is yet another normalization factor, 
\begin{equation}
\label{t-prod}
T\equiv e^{-\frac{1}{2}\frac{\delta}{\delta\phi}D\frac{\delta}{\delta\phi}}
\end{equation}
is the $T$-product, and $A(\phi)\equiv \bar{W}(-iD^{-1}\phi)$ is the generating functional of connected amplitudes with more than two external legs.

We expressed the bare action in terms of connected amplitudes and exact propagator:
\begin{equation}
\label{action}
-S(\phi)=\log{\bar{N}} -\frac{1}{2}\phi D^{-1}\phi + 
\log{T e^{A(\phi)}}.
\end{equation}
The last term in the right hand side, if expanded in powers of $A$, generates connected Feynman amplitudes whose propagators are 
$(-D)$, and vertexes, $A(\phi)$. This is a form of the Dominicis-Englert duality \cite{deDominicis:1967px}.

We now define a projector $P$ acting in the linear space of functionals of the fields. Let the arbitrary functional belonging to this space be
\begin{equation}
\label{F}
F(\phi)=\sum_{n\geq 0}\int\,dx_1\dots dx_nF_n(x_1,\dots,x_n)\phi(x_1)\dots\phi(x_n).
\end{equation}
Then $P$ is defined as follows:
\begin{equation}
\label{P}
PF(\phi)=\sum_{n >4 }\int\,dx_1\dots dx_nF_n(x_1,\dots,x_n)\phi(x_1)\dots\phi(x_n).
\end{equation}
This means that $P$ drops the first five terms in the expansion of $F$ in powers of $\phi$.

In this paper, we consider bare actions that are polynomials in the fields of the fourth power. That is $PS(\phi)=0$. Thus, acting on both sides of Eq. (\ref{action}) with $P$ gives the following inaction
equation:
\begin{equation}
\label{i-eq}
P\log{T e^{A(\phi)}}=0.
\end{equation}

It will be convenient to further transform this equation. To make this transformation, we notice that
\begin{equation}
\label{invariance}
PT^{-1}(1-P)=0.
\end{equation}
Here 
\begin{equation}
\label{inv-t}
T^{-1}\equiv e^{\frac{1}{2}\frac{\delta}{\delta\phi}D\frac{\delta}{\delta\phi}},
\end{equation}
and indeed $T^{-1}T=TT^{-1}=1$. The meaning of this equation is that both $T$-product and the inverse $T$-product $T^{-1}$ keep the kernel of $P$ intact. Indeed, the kernel of $P$ consists of the field polynomials of the fourth power, and any polynomial of the fourth power is transformed to another polynomial of the fourth power by the action of $T$ and $T^{-1}$.
Eq. (\ref{invariance}) implies that
\begin{equation}
\label{inv}
PT^{-1}(1-P)\log{T e^{A(\phi)}}=0.
\end{equation}

Now, adding the last equation to Eq. (\ref{i-eq}) we obtain the desired form of the inaction equation:
\begin{equation}
\label{in-eq-1}
PT^{-1}\log{T e^{A(\phi)}}=0.
\end{equation}
This equation is equivalent to Eq. (\ref{i-eq}), but it is preferable to the latter equation because in the linear approximation in $A$ its left hand side is just $PA$ instead of $PTA$ of Eq. (\ref{i-eq}).

To understand the character of the restriction Eq. (\ref{i-eq}) imposes on $A(\phi)$, it is useful to consider an analogous equation in a finite-dimentional case. So, we assume for a moment that $A$ is a vector of a finite-dimensional linear space, and consider an equation
\begin{equation}
\label{analog}
PA=\Phi(A),
\end{equation}
where $P$ is any projector acting in our analog finite-dimensional space, and $\Phi$ is any smooth vector-valued mapping of our space to the range of $P$ such that $\Phi(A)\approx 0$ in the linear approxiamation at small $A$.

In this finite-dimentional analog of our situation, by the implicit function theorem, Eq. (\ref{analog}) suffices to determine the part of $A$ in the range of $P$, $V\equiv PA$, as a function of the part of $A$ in the kernel of $P$, $I\equiv A-V$, at least at small $I$:
\begin{equation}
\label{implicit}
V=\mathcal{Q}(I).
\end{equation}

The mapping $\mathcal{Q}$ maps the kernel of $P$ to its range, and can be determined from the recursive eqution
\begin{equation}
\label{recursion}
V=\Phi(I+V),
\end{equation}
which is implied by Eq. (\ref{analog}) and the definitions of $I$ and $V$.

Now, using this finite-dimensional analogy, we return to our case: $A$ is the generating functional of connected amplitudes; 
$I\equiv (1-P)A$ is the inaction functional (the generating functional of connected amplitudes with not more than four particles involved); $V\equiv PA$ is the vertex functional (the generating functional of connected amplitudes with more than four particles involved). Furthermore, $V=\mathcal{Q}[I]$ expresses $V$ in terms of $I$ (the square brackets are to recall that the argument of $\mathcal{Q}$ is now a functional). We call $\mathcal{Q}$ the quantum transform. Practically, $\mathcal{Q}$ is determined from the recursive equation 
\begin{equation}
\label{recursion-1}
V=\Phi[I+V],
\end{equation}
where
\begin{equation}
\label{Phi}
\Phi\equiv P(1-T^{-1}\log T\exp).
\end{equation}

To clarify the notations in Eq. (\ref{Phi}), let us write down the leading term in the expansion of $\Phi[A]$ in powers of $A$:
\begin{eqnarray}
\label{expansion}
P(1-T^{-1}\log {T\exp})[A]\nonumber\\
\approx  P\big(A-T^{-1}\log{(1+TA+\frac{1}{2}TA^2)\big)}\nonumber\\
\approx V-PT^{-1}\big(TA+\frac{1}{2}T(A^2)-\frac{1}{2}(TA)^2\big)\nonumber\\
= \frac{1}{2}PT^{-1}\big((TA)^2-TA^2\big).
\end{eqnarray}
Notice how the first two terms in the third line are canceled against each other to yield the last line.

In a particular renormalizable model, the inaction functional $I$ spans a finite dimensional surface parameterized with the renormalized couplings of the model. This \emph{physical} surface is immersed into the infinite-dimensional kernel of $P$. If we confine $I$ to this surface, and compute $V$ in the standard way in terms of the renormalized couplings, $V$ and $I$ taken at particular values of the couplings satisfy the recursive equation (\ref{recursion-1}).

Our strategy is to consider the recursive equation (\ref{recursion-1}) as an equation for $V$ depending parametrically on $I$. It is technically convenient to allow $I$ to vary arbitrarily inside the kernel of $P$. In this way, we will obtain $V=\mathcal{Q}[I]$ defined (at least, formally) for arbitrary $I$ belonging to the kernel of $P$. The physical results are restored by restricting the quantum transform $\mathcal{Q}$ to the physical surface. 

Our approach in this paper is perturbative: we expand $\mathcal{Q}[I]$ in powers of $I$. To this end, it is convenient to introduce an auxiliary small parameter $\lambda$ and consider the dependence of $V$ on $\lambda$ defined from the equation
\begin{equation}
\label{lambda}
V(\lambda)=\Phi[\lambda I + V(\lambda)].
\end{equation}
We take that $V(0)=0$ because it corresponds to $I=0$, which implies vanishing of the couplings, and, consequently, vanishing of $V$. (This choice is consistent because $\Phi[0]=0$.) Therefore, the expansion of $V(\lambda)$ in powers of $\lambda$ may start from a linear term. But, because the leading term in expansion of $\Phi[A]$ in powers of $A$ starts from the quadratic term, Eq. (\ref{expansion}), the first derivative in $\lambda$ of $V$ vanishes at zero $\lambda$, and the expansion of $V(\lambda)$ in powers of $\lambda$ starts from a quadratic term.

In obtaining this leading quadratic term, we can neglect $V(\lambda)$ in the right hand side of Eq. (\ref{lambda}), because it contributes starting only from the cubic term. Using Eq. (\ref{expansion}) we obtain for the leading term of the expansion of $V(\lambda)$ in powers of $\lambda$ the following expression:
\begin{equation}
\label{leading}
V_2=\frac{\lambda^2}{2}PT^{-1}\big((TI)^2-TI^2\big).
\end{equation}
Here the subscript 2 in the left hand side denotes the order in $\lambda$.

The expression $T^{-1}\big((TI)^2-TI^2\big)$ involved above in $V_2$ can be simplified as follows:
\begin{equation}
\label{simplification}
T^{-1}\big((TI)^2-TI^2\big)=T^{-1}(TI)^2-I^2.
\end{equation}
Here in the second term the operations $T^{-1}$ and $T$ have canceled against each other. 

A partial cancellation is also possible for the first term in the right hand side of Eq. (\ref{simplification}). To make it, recall the definition of $T^{-1}$-product, Eq. (\ref{inv-t}). Starting from this definition it is straightforward to demonstrate a familiar property of $T$-products:
\begin{equation}
\label{t-property}
T^{-1}A_{1}A_2=e^{l_{12}}(T^{-1}A_1)(T^{-1}A_2),
\end{equation}
where the operation $l_{12}$ generates a line \emph{joining} the vertexes $A_1$ and $A_2$:
\begin{equation}
\label{j-line}
l_{12} A_1A_2 \equiv (\frac{A_1}{\delta\phi})D(\frac{A_2}{\delta\phi}).
\end{equation}
Here a $T$-product of arbitrary functionals is expressed as a \emph{joining} $T$-product of the functionals obtained from the original ones with the $T$-product. If we visualise $A_1$ and $A_2$ as two vertexes, the above representation of a $T$-product corresponds to classifying the lines appearing in the $T$-product into three classes: starting and ending on one and the same $A_i$, and joining between them. 

Later we will use a generalization of Eq. (\ref{t-property}) to a product of more than two functionals:
\begin{equation}
\label{generalization}
T^{-1}\prod_i A_i=\exp{(\sum_{i<j}l_{ij})}\prod_i T^{-1}A_i.
\end{equation}

Returning to simplification of the right hand side of Eq. (\ref{leading}), we apply Eq. (\ref{t-property}) to the first term of the right hand side of Eq. (\ref{simplification}) to obtain that
\begin{equation}
\label{simplification-1}
V_2=\frac{\lambda^2}{2}P(e^{l_{12}}-1)I_1I_2\big|_{I_1=I_2=I}.
\end{equation}

The last step in the simplification is to expand the exponential in the right hand side in powers of $l_{12}$, take into account that $I$ is a polynomial of fields of the fourth power, and that $P$ suppresses the first four powers of the fields. Each $l_{12}$ reduces the power in fields by two. Therefore, we have that only the term linear in $l_{12}$ contributes:
\begin{equation}
\label{v2}
V_2= \frac{\lambda^2}{2}Pl_{12}I_1I_2\big|_{I_1=I_2=I}.
\end{equation}
Visually this amplitude can be represented as a graph with two vertexes $I$ and a single edge $l_{12}$ joining them. Notice that this graph contributes with the weight $1/2$, which equals the standard symmetry factor for this graph. The latter is the inverse of the order of the automorphism group of this graph, which includes the only nontrivial element permuting the two vertexes with one another \cite{Ticciati:1999qp}. 

With $V_2$ known we can compute $V_3$.To this end, we can replace $V(\lambda)$ with $V_2$ in the right hand side of Eq. (\ref{lambda}) and retain the term cubic in $\lambda$. The result will be representable as a sum of graphs with three vertexes $I$ and less than four edges $D$.

It can be shown by induction that $V_{n\geq 2}$ is representable as a sum of graphs of $n$ vertexes $I$ and less than $2(n-1)$ edges.

 The combinatorial weights with which the graphs appear in $V_n$ are to be determined. The next section describes the rules for computing them.
 
 In what follows, we do not show explicitly the auxiliary parameter $\lambda$. For example, Eq. (\ref{v2}) will be written as $V_2=\big(l_{12}I_1I_2\big|_{I_1=I_2=I}\big)/2$. 

\section{The Combinatorial Weights}

We explained above that the recursive equation (\ref{recursion-1}) can be used to determine $V$ as a power series in $I$, and the expansion starts from the second power. Here we give the rules to compute the terms of this expansion, 
\begin{equation}
\label{expansion-1}
V=\sum_{k=2}^\infty V_n,\,\,\, V_n=O(I^n).
\end{equation}

For uniformity of notation, in the following we use $V_1\equiv I$. Our next task is to determine the $n$-th order in the expansion of $\Phi[\sum_{i=1}^{\infty}V_i]$ in powers of an auxiliary parameter $\lambda$ taking that each $V_i\propto \lambda^i$.

Expanding the exponential and the logarithm in the definition of $\Phi$, and using Eq. (\ref{recursion-1}), we obtain that $V_n$ is a sum of terms, each term is a product of $V_i$ with $i$ ranging from 1 to $n-1$. Notice that the term proportional to $V_n$ does not appear in the expansion because of the presence of the identity operation in the brackets of the right hand side of Eq. (\ref{Phi}).

Each term of this expansion is marked with a double partition of $n$: first time because of the expansion of the logarithm, and the second time, of the exponential. These double partitions can be marked with lists of lists of integers representing the partitions. For example, consider a term in the expansion of $V_6$ corresponding to the partitions $t=\{\{2,2,1\},\{1\}\}$: it corresponds to the second term in the expansion of the logarithm in the powers of $(T(\exp{\sum_i V_i})-1)$, and, more specifically, to the appearance of the product $PT^{-1}[(T(V_2^2/2)V_1)(TV_1)]$ in this term. In the following, the upper level entries in the list $t$ will be called subpartitions. For example $\{2,2,1\}$ and $\{1\}$ a the two subpartitions of the above partition $t$.

\begin{figure}[h]
\centering
\includegraphics[scale=.35]{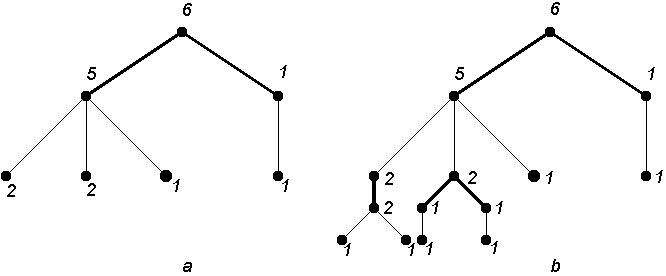}
\caption{a: The tree corresponding to the double partition $t=\{\{2,2,1\},\{1\}\}$; dark edges are related to the expansion of the logarithm and the light ones, to the expansion of the exponential of Eq. (\ref{Phi}). b: A complete tree extending the tree on the left.}
\label{tree-example}
\end{figure}

The double partition may be visualized as a two-level tree. See the tree corresponding to the above double partition on Fig. \ref{tree-example}.

A tree obtained from another tree with shuffling the branches does not give a separate contribution. To avoid double counting, we will always place ``heavier'' branches to the left from the lighter ones, like on Fig. \ref{tree-example}.

In this way we obtain
\begin{equation}
\label{two-level}
V_n=\sum_{t\in T_2[n], \,t\neq\{\{n\}\}} Z_2(t) PT^{-1}\prod_{i=1}^{L(t)}TV_{t[[i]]}.
\end{equation}
Here $n>1$, summation runs over the set $T_2[n]$ of all double partitions of $n$ excluding the partition $\{\{n\}\}$, which is the double partition of n with no actual partitions on both levels. $Z_2(t)$ is a rational coefficient we detail later (its subscript reminds that it is related to a double partition of an integer). $L(t)$ is the length of the first partition, i.e. the number of branches of the two-level tree attached to the root (for the partition from Fig. \ref{tree-example}, $L(t)=2$), and $t[[i]]$ is the $i$-th element of the partition (for the partition from Fig. \ref{tree-example}, $t[[1]]=\{2,2,1\}$, and $t[[2]]=\{1\}$).

The last unexplained object in the right hand side of Eq. (\ref{two-level}) is $V_L$ under the product, where $L$ is a list of integers. It is
\begin{equation}
\label{string}
V_L\equiv \prod_{i\in L}V_i^{p(i,L)},
\end{equation}
where the product is over the distinct elements of the list $L$, and $p(i,L)$ is the number of times the integer $i$ enters the list $L$.

The coefficient $Z_2(t)$ in Eq. (\ref{two-level}) is
\begin{equation}
\label{coefficient}
Z_2(t) = \frac{(-1)^{L(t)}(L(t)-1)!}{\prod_{s\in t}p(s,t)!}\prod_{i=1}^{L(t)}z_{t[[i]]}; \frac{1}{z_L}\equiv \prod_{i\in L}p(i,L)!.
\end{equation}
Here $L(t)$ is the length of the double partition $t$, the product in the denominator runs over the distinct subpartitions $s$ of $t$, $p(s,t)$ is the number of times $s$ enters $t$, and the notations in the definition of $z_L$ are the same as in Eq. (\ref{string}).

We now point out that both $T$-product and $T^{-1}$-product in Eq. (\ref{two-level}) generate lines of the Feynman amplitudes, and there is a partial cancellation between the amplitudes which lines come from these two $T$-products. The outcome of these partial cancellation is that the lines appear only between $V_{t[[i]]}$ and $V_{t[[j]]}$ with $i\neq j$. 

To express the above, we introduce a new operation $T_t$ depending on the tree $t$ labeling the terms of the sum in Eq. (\ref{two-level}). To introduce it, for each $V_{t[[i]]}$ under the product in Eq. (\ref{two-level}) we temporarily introduce its own set of fields $\phi_i$.
Then $T_t$ is defined as follows:
\begin{equation}
\label{new-t}
T_t\equiv\exp\big(\sum_{1\leq i<j\leq L(t)}\frac{\delta}{\delta\phi_i}D\frac{\delta}{\delta\phi_j}\big).
\end{equation}
 After application of the variational derivatives, which generate lines of the diagrams joining different $V_{t[[i]]}$, the fields are set back to the initial value, $\phi_i=\phi$. We stress that this new operation is a tool to express the fact that all the Feynman graphs appearing in the right hand side of Eq. (\ref{two-level}) and containing edges starting and ending at one and the same vertex 
$V_{t[[i]]}$ are cancelled against each other.
 
 The outcome of these considerations is the following formula for $V_n$:
 \begin{equation}
 \label{two-level-improved}
 V_n=\sum_{t\in T_2[n], \,t\neq\{\{n\}\}} Z_2(t)PT_t \prod_{i=1}^{L(t)}V_{t[[i]]}.
\end{equation}
Here the subsequent application of the two operations $T$, $T^{-1}$ in Eq. (\ref{two-level}) has been replaced with the application of the single operation $T_t$ depending on the double partition (or two-level tree) $t$.

Now we can use Eq. (\ref{two-level-improved}) recursively to determine $V_n$ in terms of $V_1\equiv I$. This is possible because there are products of $V_i$ with $i<n$ in the right hand side of this equation, and for each $V_i$ with $i>1$ we can again use the representation of Eq. (\ref{two-level-improved}).

The outcome of this recursion is the following representation:
\begin{equation}
\label{total}
V_n=\sum_{t\in T[n]} Z(t) T_t I^n.
\end{equation}
The summation here runs over the set $T[n]$ of complete trees of successive partitions of the integer $n$. By complete we mean that at the bottom of the partition there are only units, and there is no possibility of further partitioning. Each iteration adds two levels of the partition as it was above with the first iteration representing $V_n$ as a sum of products of $V_i$ with $i<n$. An example of a complete tree extending the one on the left part of Fig. \ref{tree-example} see on the right part of Fig. \ref{tree-example}. It marks a term in the sum representing $V_6$.

In Eq. (\ref{total}), the factors $Z(t)$ are the products of $Z_2(t')$ over all two-level sub-trees $t'\subset t$:
\begin{equation}
\label{factor}
Z(t)\equiv \prod_{t'\subset t, t'\in T_2} Z_2(t').
\end{equation}
For example, the complete tree on the right part of Fig. \ref{tree-example} contains three two-level sub-trees: the one on the left part of Fig. \ref{tree-example}, and two trees on a lower level representing two ways of the double partitioning of 2. Corresponding factorization of $Z(t)$ for this tree is (1/2)(-1/2)(1/2). To give a definition, a two level sub-tree $t'\subset t$ is a sub-tree of $t$ with a root on an even level of $t$ consisting of all the edges of $t$ located between the two subsequent levels and descending from the root of $t'$. (By the level we mean here the subset of the vertexes of the tree including all the vertexes at a fixed distance from the root; this distance is the number of the level; the root is the only vertex at the zero level.)

The operation $T_t$ for the case of complete trees is defined in analogy with the factor $Z_t$ above:
\begin{equation}
\label{tree-product}
T_t\equiv \prod_{t'\subset t, t'\in T_2} PT_{t'},
\end{equation}
where $T_{t'}$ is defined in Eq. (\ref{new-t}).

Each term in the sum of Eq. (\ref{total}) is a sum of Feynman amplitudes $A(g)$ corresponding to graphs $g=(V,L)$, where $V$ is the set of vertexes and $L$ is the set of edges of the graph $g$: 
\begin{equation}
\label{diagrams}
T_tI^n=\sum_{g\in G_n} C[t,g] A(g).
\end{equation}
Here $G_n$ is the set of graphs with the number of vertexes $|V|=n$, and the number of edges $|L|$ restricted in a certain way (see below). No edge of $g\in G_n$ can start and end on the same vertex. No vertex of $g\in G_n$ has more than 4 incident edges. 

The amplitudes $A(g)$ are defined as follows:
\begin{equation}
\label{amplitude}
A(g)\equiv \Big[\prod_{l\in L} \big(\frac{\delta}{\delta\phi_{l_1}}D\frac{\delta}{\delta\phi_{l_2}}\big)\prod_{v\in V} I(\phi_v)\Big]|_{\phi_v=\phi}.
\end{equation}
Here the first product runs over the edges of $g$, $l_1$ and $l_2$ are the vertexes joined by the edge $l$; the second product runs over the vertexes of $g$; the field argument of each $I$ is labeled with a vertex; after taking all the variational derivatives all the fields are set to the common value $\phi$.

Because of the involvement of the operation $P$ in the definition of $T_t$ of Eq. (\ref{total}), the expansion of $A(g)$ in powers of the fields should start from the fifth power. This is achieved by restricting the number of edges of $g\in G_n$. Indeed, the amplitude $A(g)$ of Eq. (\ref{amplitude}) is a polynomial in fields whose power is $4n-2|L|$. The concrete form of this last restriction depends on the presence of the three-point vertexes in $I$. If three point vertexes are present, expansion of $V$ in the fields starts from the fifth power, overwise it starts from the sixth power. To simplify the presentation, we assume from this point on that the inaction functional $I$ contains only four-point vertexes. In this case, the number of edges should not exceed $2n-3$. 

Substituting representation Eq. (\ref{diagrams}) in Eq. (\ref{total}) and changing the order of summation, we finally obtain
\begin{equation}
\label{finally}
V_n=\sum_{g\in G_n} C(g) A(g).
\end{equation}

The combinatorial factor $C(g)$ is defined as follows:
\begin{equation}
\label{factor}
C(g)\equiv \sum_{t\in T[n]}Z(t)C[t,g].
\end{equation}
Here the summation runs over the set $T[n]$ of all the complete trees of partitions of the integer $n$.

We now discuss the combinatorial factor $C[t,g]$ of Eq. (\ref{diagrams}). Like the conventional symmetry factor (see, e.g.,  \cite{Ticciati:1999qp}), $C[t,g]$ is inverse proportional to the size of the finite group of automorphisms of $g$, $|Aut(g)|$. But it is not proportional to $n!$, which is the number of ways to label the $n$ vertexes of $g$. Because the graphs $g$ appear in our case via the action of the operation $T_t$ depending on the complete tree $t$, not all of the possible labelings of the vertexes of $g$ contribute to $C[t,g]$. The number of the contributing labelings is $N(t,g)\leq n!$. As a result,
\begin{equation}
\label{automorphisms}
C[t,g]=\frac{N(t,g)}{|Aut(g)|},
\end{equation}  
where $N(t,g)$ is the number of contributing labelings of the graph $g$ depending on $t$.

To define $N(t,g)$ of Eq. (\ref{automorphisms}), we consider labelings of the vertexes of $g$ with the terminating vertexes of the complete tree $t$ (the number of terminating vertexes of $t$ coincides with the number of vertexes of $g$). Now we give the criteria for determination of the labelings that do not contribute to $N(t,g)$. 

As discussed above, each complete tree defines a set of two-level sub-trees. The criteria are related to this set. To check them, for each two-level sub-tree $t'\subset t$, consider its completion inside $t$, $t'_c\subset t$ ($t'_c$ consists of all the edges of $t$ descending from the root of $t'$). It defines a subgraph $g'\subset g$ as the set of vertexes labeled by the terminating vertexes of $t'_c$, and all the edges of $g$ joining these vertexes. A labeling does not contribute to $N(t,g)$ if there is a two-level sub-tree whose subgraph $g'$ contains too many lines: $|L(g')|>2 n' - 3$, where $n'$ is the number of vertexes in $g'$. This is the first criterion.

The second and the last criterion involves a two-level partition of the vertexes of the above $g'$ related to a two-level sub-tree $t'\subset t$. To define this partition, consider an edge $l\in t'$ on the first level of $t'$. It defines a subset of the vertexes $V'_l\subset V(g')$ labeled by the terminating vertexes of $t'_c$ descending from the lower end point of the edge $l$. These subsets form a partition of the vertexes of $g'$. In turn, consider an edge $e_l\in t'$ originating from the lower end point of an edge $l$. As above, it defines a subset of vertexes $V'_{e_l}\subset V'_l$. At fixed $l$, the subsets $V'_{e_l}$ form a partition of $V'_l$.  A labeling does not contribute to $N(t,g)$ if there is a two-level sub-tree containing a first level edge $l$ and second level edges $e_l\neq e'_l$ attached to $l$ such that the subgraph $g'$ contains edges joining vertexes from $V_{e_l}$ to vertexes from $V_{e'_l}$.  

Next we apply the above rules to compute the leading and the next-to-leading order of the quantum transform applied to an inaction functional not containing the three-point vertexes.

\section{The leading and next-to-leading orders}   

In the leading order of the perturbation theory under description, there are two graphs in $G_2$: with no lines and with a single line joining the two vertexes; due to the above rules, only the graph with a single line contributes, and its combinatorial weight is 1/2. This result was already obtained in Section \ref{quantum}, Eq. (\ref{v2}). It was obtained there without a use of the general rules described in the preceding Section. This agreement constitutes a check of the above rules.

\begin{figure}[h]
\centering
\includegraphics[scale=.25]{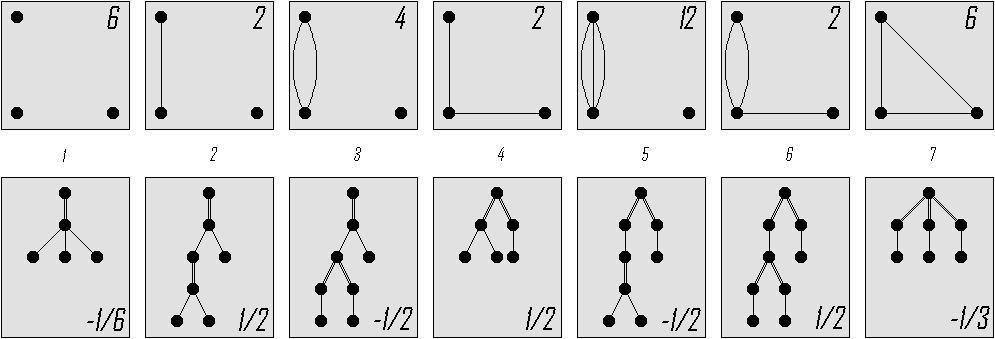}
\caption{Graphs and threes of the third order; numbers on the graph panes are $|Aut(g)|$ of Eq. (\ref{automorphisms}), on the tree panes, the $Z(t)$ factors of Eq. (\ref{factor}).}
\label{graphs-trees}
\end{figure}

In the third order, there are 7 graphs and 7 complete trees shown on the panes of Fig. \ref{graphs-trees} along with the components of the 7-dimensional ``vectors'' $|Aut(g)|$ and $Z(t)$.

$N(t,g)$ of Eq. (\ref{automorphisms}) is in this case the following 7
by 7 matrix: 
\[
N(t,g)=\left(\begin{array}{ccccccc}
6&0&0&0&0&0&0\\
6&0&0&0&0&0&0\\
6&2&0&0&0&0&0\\
6&4&4&2&2&2&0\\
6&4&4&2&2&2&0\\
6&6&4&6&4&4&6\\
6&6&6&6&6&6&6\end{array}\right).
\]
Here, for example, the first row corresponds to the tree on the first tree pane of Fig. \ref{graphs-trees}, and the first column, to the graph on the first graph pane of Fig. \ref{graphs-trees}. 

The rules for computing the combinatorial weights have been put on a computer. By now, the combinatorial weights are computed up to the sixth order of perturbation theory. This computation will be described in a separate paper. The ultimate result of this computation for $V$ can be described as follows.

$V$ is given with the following formula:
\begin{equation}
\label{result}
V= \sum_{g\in\mathcal{G}_4} \frac{A(g)}{|Aut(g)|},
\end{equation}
where the amplitude $A(g)$ is defined in Eq. (\ref{amplitude}) and $|Aut(g)|$ is the size of the group of automorphisms of the graph $g$. The summation runs over the set of graphs $\mathcal{G}_4$ defined as follows.

Any graph $g\in \mathcal{G}_4$ is connected, has the number of vertexes $|V(g)|$ and the number of edges $|L(g)|$ restricted by the inequality
$2|V(g)|-3\geq |L(g)|$. On top of that, the 
scaling dimension of any one-particle irreducible subgraph $g'\subset g\in \mathcal{G}_4$ is negative at the space-time dimension 4 
and propagator scaling dimension -2.

A scaling dimension of a one-particle irreducible subgraph $g'$ at space-time dimension $d$ and scaling dimension of the propagator -2 is $d l(g') -2 |L(g')|$, where $l(g')$ is the number of loops of $g'$ and $|L(g')|$ is its number of edges. The subscript on $\mathcal{G}_4$ is the space-time dimension. We stress that namely $\mathcal{G}_4$ appears in Eq. (\ref{result}), which singles out 4 as the preferable dimension of the space-time.

\section{Discussion}

It is remarkable in Eq. (\ref{result}) that not only the disconnected graphs have not appeared in it, but also all the connected graphs UV divergent namely in 4 space-time dimensions have completely canceled out. 

We conclude from this that a substantial part of renormalizability is of combinatoric origin. If we would start with the projector removing, say, not the five first terms in Eq. (\ref{P}), but four (corresponding to the $\phi^3$ theory), we would obtain that the preferable space-time dimension is 6. This reproduces the well known power counting conditions for renormalizable theories. 

The results obtained in the paper do not change the standard approach to renormalization of the amplitudes with four or less particles involved: one should consider convergence of the relevant Feynman integrals in this case. But beyond that, for multi-particle amplitudes, combinatorics suffices to demonstrate their UV finiteness.

We stress again that we reproduced the power counting rules from combinatorics alone, without considering Feynman integrals. 
 
 In conclusion, we have given above new rules for computing the multi-particle amplitudes as series in products of the basic amplitudes involving not more than four particles. Ultraviolet divergences do not appear in the expansion. As a by-product, we have found a combinatorial determination of the space-time dimension.
 
 The work of G.B.P. and V.T.K. was partly supported by the grants of the RF President NSh-2835.2014.2 and NSh-1633.2014.2, respectively. Also, G.B.P. thanks S.~E.~ Trunov for helpful discussions.


\end{document}